\documentclass[twocolumn]{aastex62}
\usepackage{graphicx}

\newcommand{\UDA}{Instituto de Astronom\'ia y Ciencias Planetarias, Universidad de Atacama, Copayapu 485, Copiap\'o, Chile}
\newcommand{\SAO}{Universidade de S\~ao Paulo, IAG, Rua do Mat\~ao 1226, Cidade Universit\'aria, S\~ao Paulo 05508-900, Brazil}
\newcommand{\UCN}{Instituto de Astronom\'ia, Universidad Cat\'olica del Norte, Av. Angamos 0610, Antofagasta, Chile}
\newcommand{\UDEC}{Departamento de Astronom\'\i a, Casilla 160-C, Universidad de Concepci\'on, Concepci\'on, Chile}
\newcommand{\SERENAI}{Department of Astronomy - Universidad de La Serena - Av. Juan Cisternas, 1200 North, La Serena, Chile}
\newcommand{\SERENAII}{Instituto de Investigaci\'on Multidisciplinario en Ciencia y Tecnolog\'ia, Universidad de La Serena. Benavente 980, La Serena, Chile}		
\newcommand{\UNAB}{Depto. de Cs. F\'isicas, Facultad de Ciencias Exactas, Universidad Andr\'es Bello, Av. Fern\'andez Concha 700, Las Condes, Santiago, Chile}
\newcommand{\VATICAN}{Vatican Observatory, V00120 Vatican City State, Italy}		
\newcommand{\UND}{Department of Physics and JINA Center for the Evolution of the Elements, University of Notre Dame, Notre Dame, IN 46556, USA}
\newcommand{\SUN}{School of Physics and Astronomy, Sun Yat-sen University, Zhuhai 519082, China}
\newcommand{\UNAM}{Instituto de Astronom\'ia, Universidad Nacional Aut\'onoma de M\'exico, A.P. 70-264, 04510, Ciudad de M\'exico, Mexico}
\newcommand{\VIRGINIA}{Department of Astronomy, University of Virginia, Charlottesville, VA 22904, USA}
\newcommand{\ARIZONA}{Steward Observatory, Department of Astronomy, University of Arizona, Tucson,  AZ 85721 USA}
\newcommand{\MCTI}{Observat\'orio Nacional, MCTI, Rio de Janeiro, Brazil}
\newcommand{\NOAO}{NSF's NOIRLab, 950 N. Cherry Ave. Tucson, AZ 85719 USA}
\newcommand{\ELTEI}{ELTE E\"otv\"os Lor\'and University, Gothard Astrophysical Observatory, 9700 Szombathely, Szent Imre H. st. 112, Hungary}
\newcommand{\ELTEII}{MTA-ELTE Lend{\"u}let Milky Way Research Group, Hungary}
\newcommand{\ELTEIII}{MTA-ELTE Exoplanet Research Group}

\begin{document}

\title{APOGEE-2S Discovery of Light- and Heavy-Element Abundance Correlations in the Bulge Globular Cluster NGC~6380}

\correspondingauthor{Jos\'e G. Fern\'andez-Trincado et al.}
\email{jfernandezt87@gmail.com}

\author[0000-0003-3526-5052]{Jos\'e G. Fern\'andez-Trincado}
\affil{\UCN}

\author{Timothy C. Beers}
\affil{\UND}

\author{Beatriz Barbuy}
\affil{\SAO}

\author{Szabolcs~M{\'e}sz{\'a}ros}
\affil{\ELTEI}
\affil{\ELTEII}
\affil{\ELTEIII}
	
\author{Dante Minniti}
\affil{\UNAB}
\affil{\VATICAN}

\author{Verne V. Smith}
\affil{\NOAO}

\author{Katia Cunha}
\affil{\ARIZONA}
\affil{\MCTI}

\author{Sandro Villanova}
\affil{\UDEC}

\author{Doug Geisler}
\affil{\UDEC}
\affil{\SERENAI}
\affil{\SERENAII}

\author[0000-0003-2025-3147]{Steven R. Majewski}
\affil{\VIRGINIA}

\author{Leticia Carigi}
\affil{\UNAM}

\author{Baitian Tang}
\affil{\SUN}

\author{Christian Moni Bidin}
\affil{\UCN}

\author{Katherine Vieira}
\affil{\UDA}

\begin{abstract}
 We derive abundance ratios for nine stars in the relatively high-metallicity bulge globular cluster NGC~6380. We find a mean cluster metallicity between [Fe/H]$= -0.80$ and $-0.73$, with no clear evidence for a variation in iron abundances beyond the observational errors. Stars with strongly enhanced in [N/Fe] abundance ratios populate the cluster, and are anti-correlated with [C/Fe], trends that are considered a signal of the multiple-population phenomenon in this cluster. We detect an apparent intrinsic star-to-star spread ($\gtrsim 0.27$ dex) in the slow neutron-capture process element (\textit{s}-element) Ce II. Moreover, the [Ce/Fe] abundance ratio exhibits a likely correlation with [N/Fe], and a somewhat weaker correlation with [Al/Fe]. If confirmed, NGC~6380 could be the first high-metallicity globular cluster where a N-Ce correlation is detected. Furthermore, this correlation suggests that Ce may also be an element involved in the multiple-population phenomenon. Currently, a consensus interpretation for the origin of the this apparent N-Ce correlation in high-metallicity clusters is lacking. We tentatively suggest that it could be reproduced by different channels -- low-mass asymptotic giant-branch stars in the high-metallicity regime or fast-rotating massive stars  (``spinstars"), due to the rotational mixing.  It may also be the cumulative effect of several pollution events including the occurrence of peculiar stars. Our findings should guide stellar nucleosynthesis models, in order to understand the reasons for its apparent exclusivity in relatively high-metallicity globular clusters.
\end{abstract}

%%%%%%%%%%%%%%%%%%%%%%%%%%%%%%%%%%%%%%%%%%%%%%%%%
%%%%% INTRODUCTION 
%%%%%%%%%%%%%%%%%%%%%%%%%%%%%%%%%%%%%%%%%%%%%%%%%
\section{INTRODUCTION} 
\label{section1}

For over a decade, spectroscopic and photometric data have revealed that virtually all studied Milky Way (MW) Globular Clusters (GCs) host (at least) two main groups of stars with a complex chemical-enrichment history -- the so-called multiple-population (MP) phenomenon in GCs, which are commonly disinguished by their different light-element enrichment \citep[see review by][]{Bastian2018}.

In general, GC stars have been shown to exhibit C, N, O, Mg, Na, and Al variations \citep[see, e.g.,][and references therein]{Carretta2009a, Meszaros2015, Pancino2017, Schiavon2017, Masseron2019, Nataf2019,  Meszaros2020, Geisler2021}, with a few exceptions exhibiting K and Ca abundance variations \citep{Cohen2012, Carretta2013, Carretta2021}, as well as internal variations in the heavy elements produced via slow neutron-capture reactions \citep[][]{Masseron2019, Meszaros2020, Marino2021}. In addition, the most common feature identified so far in almost all the GCs are the clear anti-correlations Na-O and N-C, Mg-K, and Al-Mg \citep[see, ][]{Mucciarelli2015, Pancino2017, Masseron2019, Meszaros2020}, some of them depending on the GC mass and metallicity \citep[see, e.g.,][]{Pancino2017, Carretta2021}, as well as correlations between N-Al, Al-Si, and others\citep[see, e.g.,][]{Masseron2019, Meszaros2020}. 

Even though a broad range of polluters \citep[see][for a review]{Renzini2015} have been proposed to explain these apparent abundance variations in GCs, a clear consensus on the origin of the nucleosynthetic pathways that were responsible for the puzzling (anti-) correlations among light elements is still under debate.

In this Letter, we analyze recent APOGEE-2S data of the heavily reddened \citep[E(B$-$V)$=$1.07; ][]{Ortolani1998} bulge GC NGC~6380 \citep{Djorgovski1993}, also known as Tonantzintla~1 and Pismis 25--\citep{Pismi1959}. 

 We suggest that a correlation between Al and N with Ce could be present in NGC~6380, as suggested by the Pearson and Spearman correlation coefficients, indicating that the multiple-population phenomenon includes this \textit{s}-process element as well as the other traditional light elements associated with it. However, additional observations are needed to confirm this finding and to gain more confidence in this assertion. This finding could play an important role as a new entry among the chemical anomalies in GCs at metallicity around [Fe/H]$=-0.73$, and will supply useful additional constraints to theoretical models to infer more appropriate scenarios for the origin of MPs in GCs.

\begin{figure*}
	\begin{center}
		\includegraphics[width=180mm]{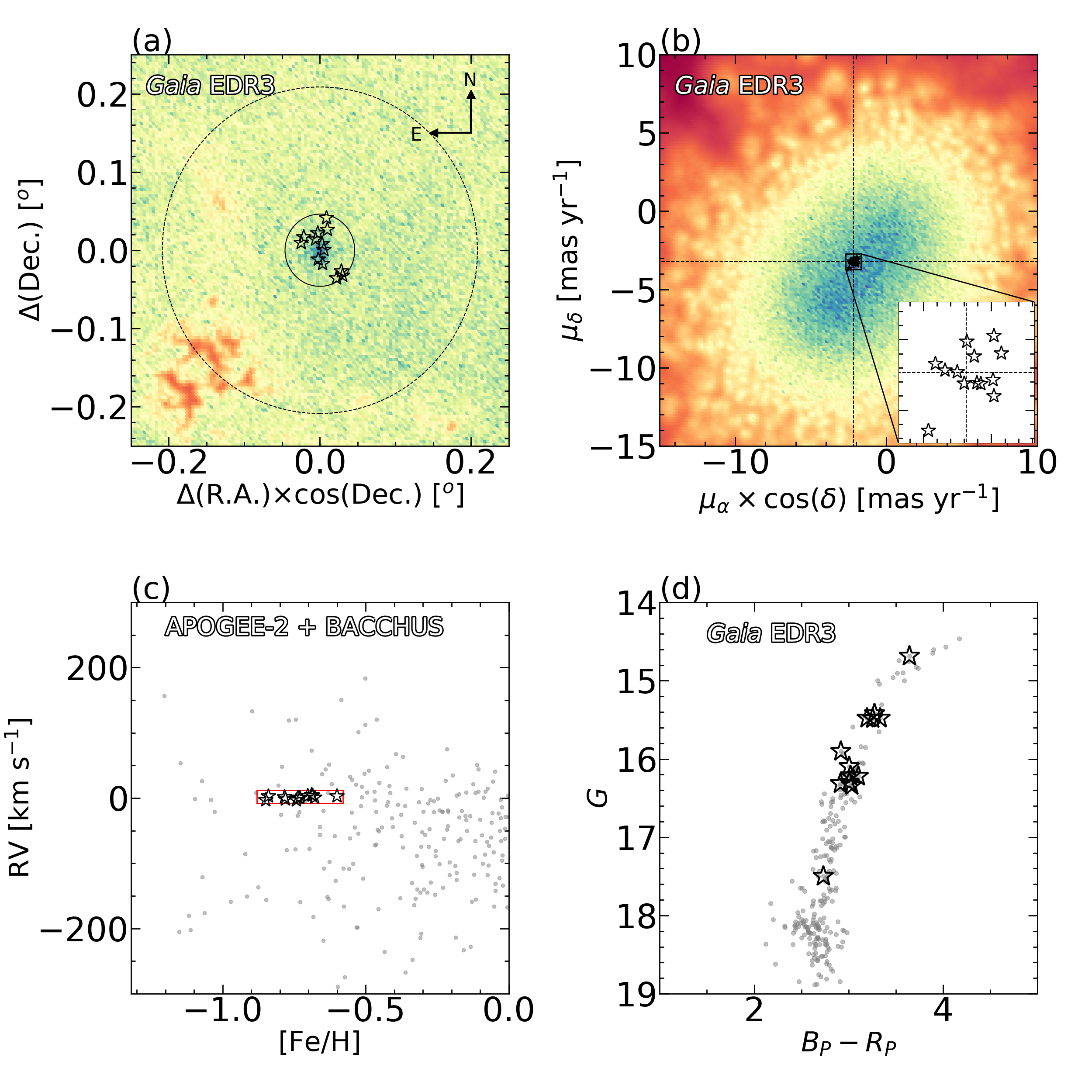}
		\caption{Main properties of NGC~6380 stars. Panel (a): Sky position of stars centered on NGC~6380. Stars analyzed in this work are marked with star symbols, while the overlaid large and small circles refer to the cluster tidal radius (${\rm r_{t}}$) and twice the half-mass radius (${\rm 2\times r_{h,m}}$), respectively. Density maps in panel (a) highlight the stars in the \textit{Gaia} EDR3 footprint within a box of 0.5$\times$0.5 $\deg$ centered on NGC~6380, while panel (b) shows the distribution in PMs for stars toward the field of NGC~6380, with the inner window highlighting the distribution of our sample within a 0.5 mas yr$^{-1}$ radius around the nominal PMs of the cluster highlighted with black dashed lines. Panel (c): Radial velocity versus [Fe/H] of our member stars compared to APOGEE-2 field stars. The metallicity of our targets have been determined with \texttt{BACCHUS}, while those for field stars are from the \texttt{ASPCAP} pipeline. The red box limited by $\pm$0.15 dex and $\pm$10 km s$^{-1}$, centered on [Fe/H] $= -0.73$ and RV $= +1.92$ km s$^{-1}$, encloses our potential cluster members. Panel (d): Differential-reddening corrected color-magnitude diagram in the \textit{Gaia} bands for cluster stars with a membership probability larger than 90\% \citep[see, e.g.,][]{Vasiliev2021}.}
		\label{Figure1}
	\end{center}
\end{figure*}

\section{DATA} 
\label{section2}

We use data from the sixteenth data release \citep[DR16;][]{Ahumada2020} of the second generation of the Apache Point Observatory Galactic Evolution Experiment (APOGEE-2) survey \citep{Majewski2017}, which is one of the programs in the Sloan Digital Sky Survey \citep[SDSS-IV;][]{Blanton2017}. 

The APOGEE-2 instruments are high-resolution ($R \sim 22,000$), near-infrared (collecting most of the \textit{H}-band: $15145$--$16960$ \AA{}; vacuum wavelengths) spectrographs \citep{Wilson2019} that operate on the Sloan 2.5m telescope \citep{Gunn2006} at Apache Point Observatory (APOGEE-2N) and on the Ir\'en\'ee du Pont 2.5m telescope \citep{Bowen1973} at Las Campanas Observatory (APOGEE-2S). The targeting strategy of the APOGEE-2 survey is summarized in \citet{Zasowski2017}, while spectra are reduced as described in \citet{Nidever2015}, and analyzed using the APOGEE Stellar Parameters and Chemical Abundance Pipeline \citep[\texttt{ASPCAP};][]{Garcia2016}.  The libraries of synthetic spectra and the \textit{H}-band line list used are described in \citet{Zamora2015} and \citep{Smith2021}, respectively. 

\begin{figure}
	\begin{center}
		\includegraphics[width=80mm]{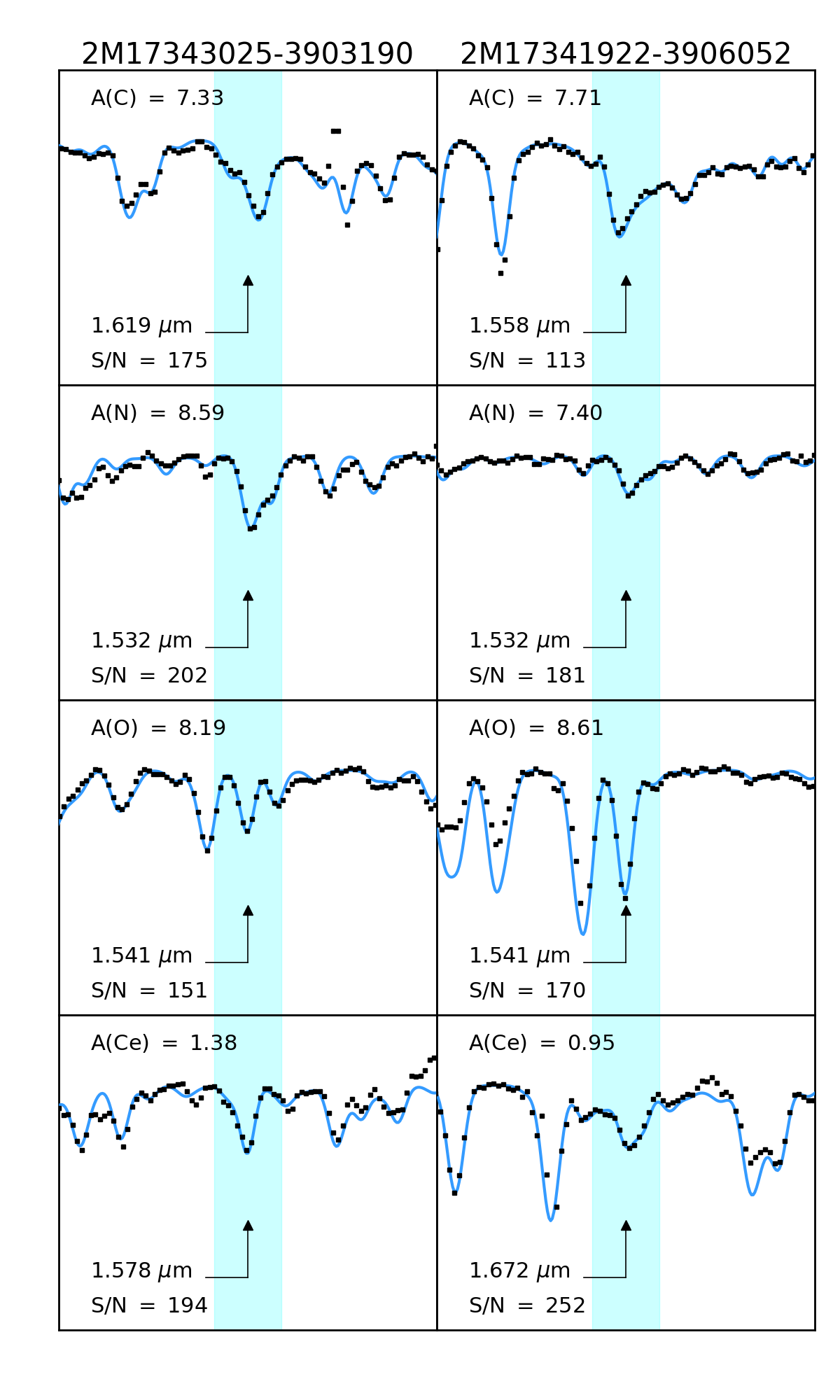}
		\caption{Example on the detection of $^{12}$C$^{16}$O, $^{16}$OH, $^{12}$C$^{14}$N, and Ce II lines in two arbitrarily selected stars in our sample. A portion of the spectral synthesis is shown for the determination of the [C/Fe], [N/Fe], [O/Fe], and [Ce/Fe] abundances for two stars in the innermost regions of NGC~6380, and by adopting the spectroscopic atmospheric parameters listed in Table \ref{Table1}. Each panel shows the best-fit synthesis (blue) from \texttt{BACCHUS} compared to the observed spectra (black squares) of selected lines (marked with black arrows, and cyan shadow bands of 3.2$\times 10^{-4}$ $\mu$m wide).
		}
		\label{Figure2}
	\end{center}
\end{figure}

\subsection{NGC~6380} 
\label{section3}

The APOGEE-2S plug-plate containing the NGC~6380 stars was centered on $(l,b)\sim (350 ^{\circ},-04 ^{\circ})$ as part of the bulge program survey, and 13 of 506 science fibers were located in the innermost region ($\lesssim 2\times {\rm r_{h,m}}$) of NGC~6380, as shown in Figure \ref{Figure1}(a). 

Figures \ref{Figure1}(b--d) reveal that our stars share the same kinematic and astrometric properties as NGC~6380 (also listed in Table \ref{Table1}), and are positioned along the red giant branch (RGB) of the cluster. The nominal proper motions of the cluster, highlighted in Figure \ref{Figure1}(b), have been taken from \citet{Vasiliev2021}-- ($\mu_{\alpha}\cos{}(\delta)$, $\mu_{\delta}$)$=$($-2.183\pm0.031$ mas yr$^{-1}$,$-3.233\pm0.03$ mas yr$^{-1}$), while the structural parameters of the cluster are taken from \citet{Cohen2021} -- (${\rm r_{h,m}}$, ${\rm r_{t}}$)$=$(83", 751").

It is important to note that 12 of the 13 observed cluster stars in our sample have spectra with a high signal-to-noise (S/N), $>75$ pixel$^{-1}$, except for one star, 2M17342736$-$3905102, which has a spectrum with a S/N$= 29$ pixel$^{-1}$, and is the faintest star in Figure \ref{Figure1}(d). 

In the following, we use all stars to provide reliable and precise ($<$ 1 km s$^{-1}$) radial velocities for cluster-member confirmation, but limit ourselves to the 12 higher S/N stars for the abundance analysis, as highlighted in Figure \ref{Figure1}(c). It is also important to note that other sources that fall inside the box highlighted in Figure \ref{Figure1}(c) are foreground/background stars with other properties that are not compatible with NGC~6380 stars. 

\begin{figure*}
	\begin{center}
		\includegraphics[width=180mm]{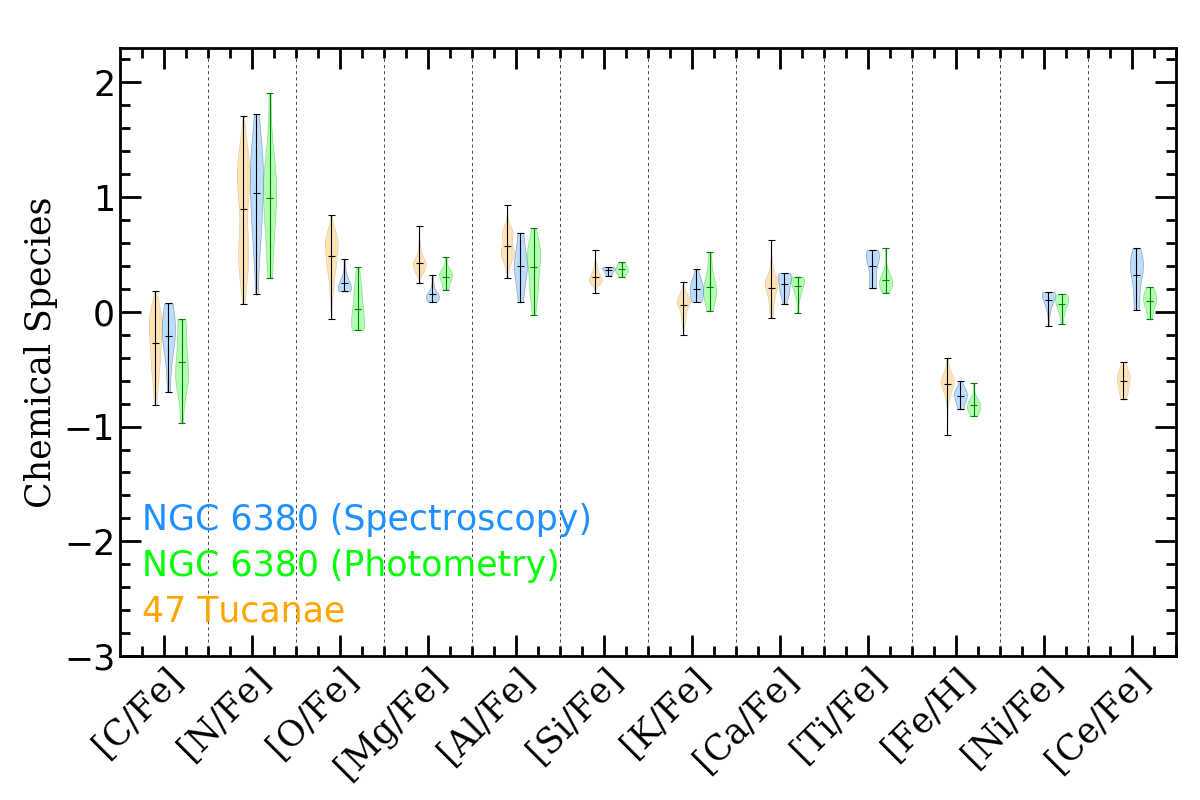}
		\caption{Elemental-abundance density estimation (violin representation) of NGC~6380 (dodgerblue), compared to elemental abundances of 47 Tucanae (orange) from \citet{Meszaros2020}. Each violin representation indicates with lines the mean and limits of the distribution.}
		\label{Figure3}
	\end{center}
\end{figure*}

\section{Elemental abundances} 
\label{section4}

Atmospheric parameters were adopted from the DR16 (uncalibrated) T$_{\rm eff}$ and $\log$ \textit{g} determined by the \texttt{ASPCAP} pipeline through the best-fits between the synthetic and observed spectra for the entire APOGEE region; and also by applying a simple approach of fixing T$_{\rm eff}$ and log \textit{g} to values determined independently of spectroscopy. We obtain T$_{\rm eff}$ and $\log$ \textit{g} from photometry in the same manner as described in \citet{Romero-Colmenares2021}, i.e., we first derived the differentially reddening-corrected color magnitude diagram (CMD) of Figure \ref{Figure1}(d). We then horizontally projected the position of each observed star until it intersected the \texttt{PARSEC} \citep{Bressan2012} isochrone (chosen to be 12 Gyr), and assumed T$_{\rm eff}$ and $\log$ \textit{g} to be the temperature and gravity of the point of the isochrones that have the same G magnitude as the star. We underline the fact that, for highly reddened objects like NGC~6380, the absorption correction depends on the spectral energy distribution of the star, i.e., on its temperature. For this reason, we applied a temperature-dependent absortion correction to the isochrone. Without this, it is not possible to obtain a proper fit of the RGB, especially of the upper and cooler part. The adopted atmospheric parameters are listed in Table \ref{Table1}. It is important to note that the target stars are not compatible with both the set of parameters. This issue does not strongly affect the mean derived [X/Fe] abundance ratios in NGC~6380, but some chemical species, such as oxygen, magnesium, and cerium, are more sensitive to these discrepancies, as can be appreciated in Table \ref{Table1}. We highlight that our determined abundance ratios by adopting atmospheric parameters from photometry are likely more reliable and realistic, as the set of photometry parameters better reflect the real distribution of the stars on the CMD.
	
With the fixed T$_{\rm eff}$ and $\log$ \textit{g}, the first step consisted in determining the metallicity from selected Fe I lines, the micro-turbulence velocity ($\xi_{t}$), and the convolution parameter with the Brussels Automatic Stellar Parameter (\texttt{BACCHUS}) code \citep{Masseron2016}. Thus, the metallicity provided is the average abundance of selected Fe I lines, while the micro-turbulence velocity is obtained by minimizing the trend of Fe abundances against their reduced equivalent width, and the convolution parameter represents the total effect of the instrument resolution.

Chemical abundances were derived from a local thermodynamic equilibrium (LTE) analysis using the \texttt{BACCHUS} code and the \texttt{MARCS} model atmospheres \citep{Gustafsson2008} for likely NGC~6380 stars, following the same methodology as described in \citet{Fernandez-Trincado2019, Fernandez-Trincado2020, Fernandez-Trincado2021a, Fernandez-Trincado2021b, Fernandez-Trincado2021c}. This approach allow us to minimize the number of caveats present in the determinations of \texttt{ASPCAP} abundance ratios toward GCs \citep[see, e.g.,][]{Masseron2019, Meszaros2020}. For APOGEE-2 spectra, we infer chemical abundances for the Fe-peak (Fe, Ni), odd-Z (Al, K), $\alpha$- (O, Mg, Si, Ca, Ti), light- (C, N), and \textit{s}-process (Ce) elements.

Figure \ref{Figure2} show two examples of the excellent quality of the APOGEE-2 spectra and the success of the  fitting procedure for selected atomic and molecular lines for two arbitrarily selected members of NGC~6380. The same figure also indicates the detectable $^{12}$C$^{16}$O, $^{16}$OH, $^{12}$C$^{14}$N, and Ce II lines (cyan bands in Figure \ref{Figure2}).
	
The resulting elemental-abundance ratios are listed in Table \ref{Table1}, which are scaled to the Solar reference values from \citet{Asplund2005}.

\section{Results and discussion} 
\label{section5}
%%%%%%%%%%%%%%

NGC~6380 has been examined previously in \citet{Horta2020} using \texttt{ASPCAP}/APOGEE DR16 results.  In that study they investigated if GCs formed in situ or from external origin. Based on the calibrated \texttt{ASPCAP} [Si/Fe] abundance ratios, they concluded that NGC 6380 formed in situ and did not have an extragalactic origin. However, it is important to note that their \texttt{ASPCAP} determinations of [Si/Fe] for NGC~6380 are $\sim$0.15  dex lower, on average, than our determinations. This discrepancy could be due to some issues with the accuracy (zero-point) of \texttt{ASPCAP} abundances toward GCs, limits of the model grid, and/or the difficulty of fitting lines where the intensity is comparable to the variance. Furthermore, the \texttt{ASPCAP} pipeline uses a global fit to the continuum in the three detector chips independently, while the \texttt{BACCHUS} pipeline places the pseudo-continuum in a region around the lines of interest. Thus, we believe that our manual method is likely more reliable, since it avoids possible shifts in the continuum location due to imperfections in the spectral subtraction along the full spectral range \citep[see, e.g.,][for details]{Masseron2019}.

Figure \ref{Figure3} summarizes the chemical makeup for likely members of NGC~6380 in a violin representation, which is compared to the elemental abundances of 47 Tucanae stars from high-quality APOGEE data taken from \citet[][]{Meszaros2020}. 

 We find a mean metallicity $\langle$[Fe/H]$\rangle$ between $-0.80$ and $-0.73$ for NGC~6380 depending on the adopted atmospheric parameters; within the uncertainties we do not detect any real spread in this element. Table \ref{Table1} indicates that [Fe/H] is systematically shifted within the internal uncertainties due to the adoption of a different T$_{\rm eff}$ and $\log$ \textit{g} scales. However, given the small sample size of this work, the presence of an iron spread cannot be definitively ruled out, and will be investigated in future CAPOS papers \citep{Geisler2021}.

\begin{figure*}
	\begin{center}
		\includegraphics[width=180mm]{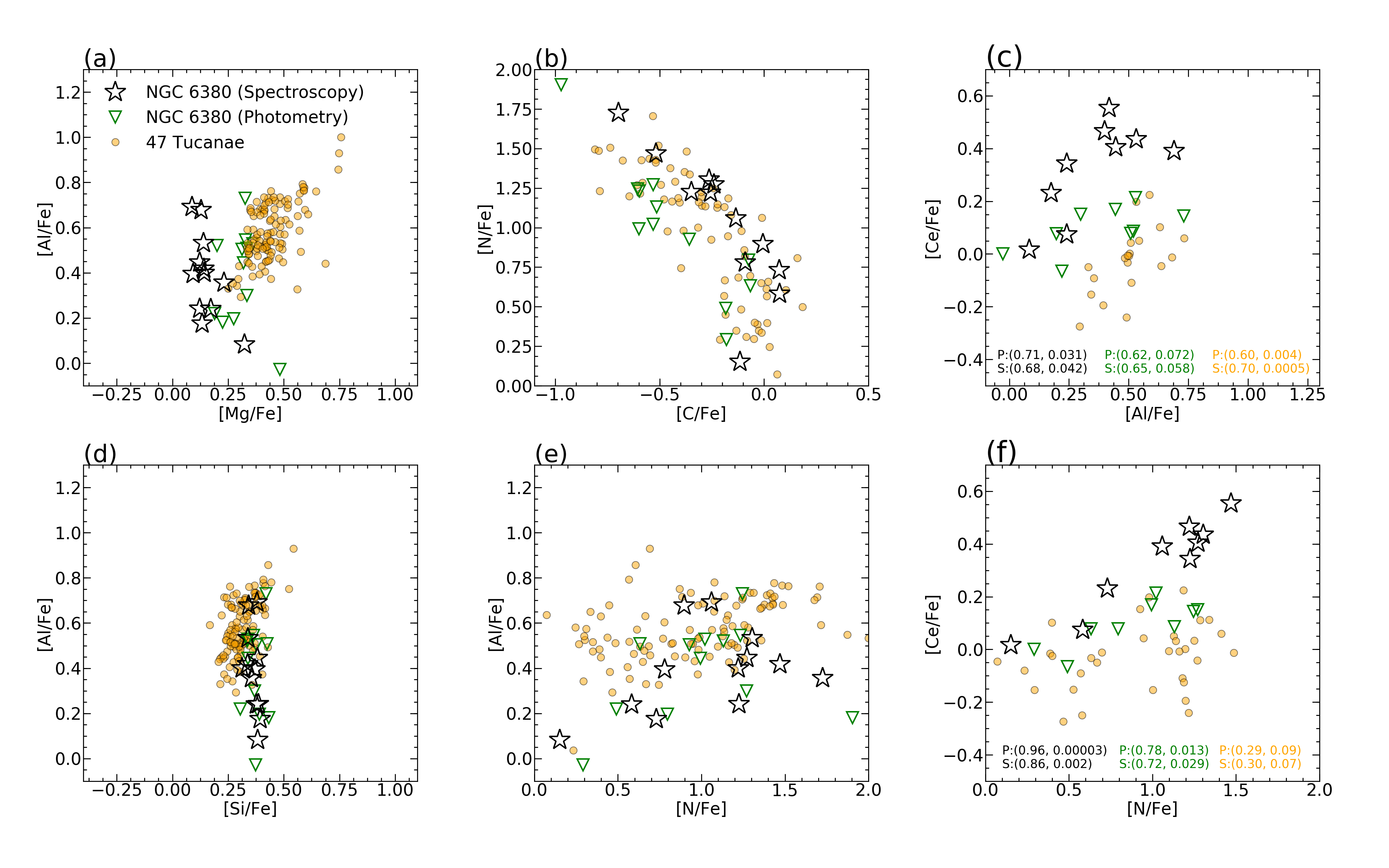}
		\caption{Panel (a)--(f):  [Mg/Fe] -- [Al/Fe], [C/Fe] -- [N/Fe], [Al/Fe]--[Ce/Fe], [Si/Fe]--[Al/Fe], [N/Fe]--[Al/Fe], and [N/Fe]--[Ce/Fe] distributions for NGC~6380 (black stars) and 47~Tucanae (orange dots) stars. In panels (c) and (f), Pearson's (P: first row of the annotation) and Spearman's (S: second row of the annotation) coefficients (first entry) and p$-$values (second entry) are indicated for our determinations by adopting atmospheric parameters from spectroscopy (black annotation), photometry (green annotation), and 47 Tucanae (orange annotation). The typical uncertainties are listed in Table \ref{Table1}.
		}
		\label{Figure4}
	\end{center}
\end{figure*}

Overall, the chemical enrichment of NGC~6380 is in agreement with 47 Tucanae, whose metallicity is very similar to that of NGC 6380, within the observational errors, for almost all species except [Ce/Fe] (see Figure \ref{Figure2}), which displays enhancements above solar in NGC~6380, but below solar in 47~Tucanae. In addition, NGC~6380 exhibits a slightly larger spread ($\gtrsim 0.27$ dex) in [Ce/Fe] that exceeds the typical observational uncertainties, while that of 47~Tucanae is smaller, as can be appreciated from inspection of Figure \ref{Figure4}(c,f) and the results listed in Table \ref{Table1}. 

Unfortunately, \citet[][]{Meszaros2020} do not provide determinations for [Ti/Fe] and [Ni/Fe] for 47 Tucanae, however, our determinations (see mean values in Table \ref{Table1}) reveal that both chemical species in NGC~6380 are at the same typical levels as other MW GCs or field stars with similar metallicity \citep[see, e.g.,][]{Villanova2019}. This is also supported by the super-solar abundances of other  $\alpha$-elements (O, Mg, Si, and Ca), with a small star-to-star spread (not significantly larger than the typical error bars), being compatible with other GCs at similar metallicity such as 47 Tucanae (Figure \ref{Figure3}). We conclude that the typical [Ti/Fe] and [Si/Fe] enrichment observed in NGC~6380, accompanied by its bulge-like orbit \citep[][]{Massari2019}, suggests it may have formed in situ. 

The odd-Z elements (Al, K) are also slightly over-abundant compared to the Sun, with $1\sigma$-spreads between $0.08$ and $0.11$ dex in K, and $\sim0.17$ dex in Al. Neither the Mg-K and Mg-Al anti-correlation nor the Si-Al and N-Al correlation are evident in NGC~6380. However, we notice that the mean $<$[Mg/Fe]$>$ is between $+0.15$ and $+0.30$ (depending on the adopted atmospheric parameters), and is less abundant than that observed in 47 Tucanae at similar metallicity (see Figures \ref{Figure3} and \ref{Figure4}), which could be related somehow with the large spread in Al  observed in NGC~6380. However, it is important to note that, if the GC polluters are asymptotic giant branch (AGB) stars, then the observed large spread in Al seems to be at odds with the predicted yields of massive AGB stars, as models suggest a modest Mg-Al cycle with no Al enrichment above [Fe/H] $> -1.0$ \citep[see e.g.,][]{Ventura2016, Meszaros2020}. 

Regarding the light-elements (C, N), we find a very large variation ($\gtrsim1.5$ dex) in N, which is clearly anti-correlated with C, and comparable to the extended variations in N and C observed in 47 Tucanae (see Figure \ref{Figure4}(b)), giving support to the idea that NGC~6380 indeed has multiple populations based on N. Indeed, at the metallicity of NGC~6380, it is likely that only second-generation stars attain such high nitrogen abundances \citep[see, e.g.,][]{Schiavon2017, Meszaros2020}. Therefore, we conclude that NGC~6380 hosts MPs.  

 Figure \ref{Figure4}(c) also shows that Ce is likely correlated with Al, albeit with a weaker amplitude, while Figure \ref{Figure4}(f) exhibits an apparent correlation with the N abundance.  The Pearson and Spearman correlation test between Ce and N is larger, with a p-value near zero, as indicated in the internal legend in Figure \ref{Figure4}(c:f), indicating that the observed correlation is not due to random chance. It is important to notice that the adoption of a different T$_{\rm eff}$ and $\log$ \textit{g} scale produces a slightly less pronounced -- but still present -- correlation (as can be appreciated in Figure \ref{Figure4}(c:f)), indicating an apparent correlation between Ce and the light elements. Additionally, as can be appreciated in Table \ref{Table1}, the observed star-to-star scatter in [Ce/Fe] is larger ($\gtrsim$0.27 dex) than the typical uncertainties in [Ce/Fe] in individual stars. However, it is also important to note that the apparent correlation between Ce, N, and Al could be affected by (systematic and random) uncertainties, due, for instance, to the differential-reddening correlation and the adopted isochrones. Therefore, it is highly desirable to obtain additional observations to confirm this finding.

In addition to the N-Ce and Al-Ce correlations seen in NGC~6380 found in this work, we examined the APOGEE-2 GC data compiled by \citet{Meszaros2020}, and found no significant correlation with Ce for other GCs at the APOGEE-2 database at a similar metallicity as NGC~6380, with the exception of 47 Tucanae, which exhibits a marginal correlation of Al and N with Ce, as shown in Figure \ref{Figure4}(c:f). It is important to note that NGC~6380 (3.34$\times$10$^{5}$ M$_{\odot}$) is less massive than 47~Tucanae (8.95$\times$10$^{5}$ M$_{\odot}$) according to the recent estimation by \citet{Baumgardt2018}, likely indicating a GCs mass threshold for the ocurrence of N-Ce or Al-Ce correlations.

 NGC~6380 is \textit{the first case of a bulge GC where a N-Ce correlation has been detected}, likely indicating a different chemical evolution of this cluster with respect to the bulk of the MW GCs. Currently, a consensus interpretation of the origin for such N-Ce or Al-Ce correlations is still lacking.

The origins of the cerium abundance ratios and enhanced nitrogen are possibly related to AGB stars, as recently reconfirmed by \citet{Kobayashi2020}. Cerium is produced primarily via the \textit{s}-process in thermally-pulsing (TP) AGB stars, while large amounts of primary nitrogen can be synthesized as a result of hot bottom burning (HBB) in intermediate-mass AGB stars, where primary $^{12}$C from triple-$\alpha$ burning is converted to $^{14}$N in the deep convective envelope \citep[e.g.,][]{Karakas2014}. This could be an important indication that, in this cluster, the second-generation stars could be the product of yields from low- or intermediate-mass AGB stars  \citep{Ventura2009}. It the relatively high metallicity of NGC~6380, the apparent increase of the Ce abundance as N increases could support this assertion. Another possible alternative could be fast-rotating massive stars (``spinstars"), which can also produce these secondary elements as primaries, due to the rotational mixing. These effects are mainly studied in metal-poor stars,  but they were extended to different metallicities in \citet{Frischknecht2016} and \citet{Limongi2018}.

In conclusion, it is not presently possible to decide about the origin of the N and Al correlations with the \textit{s}-process element Ce, but clearly the data shows that they are correlated, since that is the expectation for production in these two possible sites (AGBs and spinstars). Further discussion regarding pollution scenarios require detailed computations and analysis, beyond the scope of the present work.

\section{Conclusions} 
\label{section6}

APOGEE-2S has enabled the first detailed study of NGC~6380 cluster, which was not possible before given the high obscuration, therefore no other determinations have ever been done from spectroscopy. Here, we report on the apparent discovery of a bulge GC (NGC~6380), at relatively high metallicity, [Fe/H] $\sim -0.80$ to $-0.73$ (depending on the adopted atmospheric parameters), where Ce is likely correlated with the N and Al abundances. By adopting different scales in the atmospheric parameters, we find systematic changes in some of the determined abundances ratios, yet the correlation between Ce with N and Al appear to remain.

The present analysis suggest two main results: (1) slow neutron-capture element variation is likely present in the bulge GC NGC~6380. It is therefore worthwhile to explore if this variation exists in other GCs, particularly in the relatively high-metallicity regime; and (2) among the MPs in this GC, the Ce abundances appear to be correlated with those of N and Al, thus Ce could be also a multiple-population element. Different processes could be responsible for the puzzling N-Ce and Al-Ce correlations, likely indicating a different chemical-evolution history for bulge GCs at metallicities similar to NGC~6380 with respect to the bulk of MW GCs.

 \floattable
 \begin{deluxetable*}{ccccccccccccccccccccccc}	
 	\center
 	\setlength{\tabcolsep}{0.5mm}  
 	\tabletypesize{\tiny}
 	\tablecolumns{23}
 	%	\tablewidth{0.pt}
 	\rotate
 	\tablecaption{Main Kinematics, Astrometric Properties, and Elemental Abundances for Likely  NGC~6380 Members}
 	\tablehead{
 		\colhead{APOGEE$-$Id}	                                  &  
 		\colhead{$G$}		                                                          &  
 		\colhead{${B}_{P}$}		                             &    
 		\colhead{${R}_{P}$}		                             &        
 		\colhead{S/N}		                                                   & 
 		\colhead{RV}		                                                    & 
 		\colhead{$\mu_{\alpha}\cos(\delta)$}     & 
 		\colhead{$\mu_{\delta}$}		                       &  		 		
 		\colhead{${\rm T_{eff}}$}	                             &  		
 		\colhead{log \textit{g}}	                                  &  		
 		\colhead{ $\xi_{t}$}		                                    & 
 		\colhead{[C/Fe]}		    &        
 		\colhead{[N/Fe]}		    & 
 		\colhead{[O/Fe]}		    & 
 		\colhead{[Mg/Fe]}		    &   
 		\colhead{[Al/Fe]}		    &  
 		\colhead{[Si/Fe]}		    &    
 		\colhead{[K/Fe]}		    &        
 		\colhead{[Ca/Fe]}		    & 
 		\colhead{[Ti/Fe]}		    & 
 		\colhead{[Fe/H]}		    & 
 		\colhead{[Ni/Fe]}		    & 		
 		\colhead{[Ce/Fe]}		\\
 		\colhead{}		    &  
 		\colhead{}		    &     
 		\colhead{}		    &  
 		\colhead{}		    &  
 		\colhead{pixel$^{-1}$}		    &  
 		\colhead{km s$^{-1}$}		    &     
 		\colhead{mas yr$^{-1}$}		    &  
 		\colhead{mas yr$^{-1}$}		    &  
 		\colhead{K}		    &  
 		\colhead{cgs}		    &  
 		\colhead{km s$^{-1}$}		    &  
 		\colhead{}		    &  
 		\colhead{}		    &   
 		\colhead{}		    &     
 		\colhead{}		    &  
 		\colhead{}		    &  
 		\colhead{}		    &  
 		\colhead{}		    &     
 		\colhead{}		    &  
 		\colhead{}		    &  
 		\colhead{}		    &  
 		\colhead{}		    &             
 		\colhead{}		                           
 	}
 	\startdata
\hline 	
{\bf Photometry} & & & & & & & & & & & & & & & & & & & & \\
\hline
  2M17342588$-$3901406 & 14.69  &  16.99   &  13.35   &  234     & $+$3.63       &  $-$2.25   &   $-$3.22 & 3584 &  0.26 & 2.37 & $-$0.18 & $+$0.49 &  $+$0.17  &  $+$0.18 & $+$0.22 &  $+$0.30 &  $+$0.39  &  $+$0.13 &  $+$0.19  & $-$0.78 & $+$0.02 &  $-$0.06 \\
2M17341922$-$3906052 & 15.42  &  17.41   &  14.14   &  160     & $+$3.73       &  $-$2.12   &   $-$3.11 & 3866 &  0.77 &  1.56 & $-$0.18 & $+$0.29 &  $+$0.39  &  $+$0.48 & $-$0.02 &  $+$0.37 &  $+$0.18  &  $+$0.29 &  $+$0.20  & $-$0.61 & $-$0.10 &  $+$0.01 \\
2M17342921$-$3904514 & 15.49  &  17.38   &  14.12   &  158     & $+$2.16       &  $-$1.98   &   $-$3.39& 3850 &  0.74 &  2.50 & $-$0.51 & $+$1.13 &  $-$0.10  &  $+$0.20 & $+$0.52 &  $+$0.33 &  $+$0.51  &  $+$0.23 &  $+$0.35  & $-$0.79 & $+$0.10 &  $+$0.08 \\
2M17343616$-$3903344 &  15.48  &  17.41   &  14.22   &  145     & $-$1.28       &  $-$2.34   &   $-$3.21 & 3890 &  0.81 & 2.31 & $-$0.53 & $+$1.27 &  $+$0.14  &  $+$0.33 & $+$0.29 &  $+$0.36 &  $+$0.27  &  $+$0.21 &  $+$0.31  & $-$0.85 & $+$0.07 &  $+$0.15 \\
2M17343025$-$3903190 & 15.48  &  17.49   &  14.16   &  144     & $+$0.63       &  $-$2.10   &   $-$3.30   & 3859 &  0.75 & 2.48 & $-$0.59 & $+$0.99 &  $-$0.09  &  $+$0.31 & $+$0.44 &  $+$0.33 &  $+$0.28  &  $+$0.28 &  $+$0.25  & $-$0.85 & $+$0.15 &  $+$0.16 \\
2M17342693$-$3904060 & 16.22  &  17.79   &  14.69   &  105     & $-$2.79       &  $-$2.46   &   $-$3.64   & 4073 &  1.14 & 2.21 & $-$0.07 & $+$0.79 &  $+$0.21  &  $+$0.27 & $+$0.19 &  $+$0.39 &  $+$0.13  &  $+$0.14 &  $+$0.16  & $-$0.90 & $+$0.13 &  $+$0.07 \\
2M17342541$-$3902338 &  15.90  &  17.60   &  14.69   &   88     & $-$2.61       &  $-$2.07   &   $-$3.31  & 4059 &  1.12 & 2.27 & $-$0.06 & $+$0.63 &  $-$0.01  &  $+$0.33 & $+$0.50 &  $+$0.42 &  $+$0.19  &  $+$0.30 &  $+$0.28  & $-$0.86 & $+$0.14 &  $+$0.07 \\
2M17342767$-$3903405 & 16.29  &  17.96   &  14.95   &   87     & $+$4.21       &  $-$2.41   &   $-$3.16 & 4125 &  1.24 & 2.10 & $-$0.53 & $+$1.02 &  $+$0.01  &  $+$0.36 & $+$0.52 &  $+$0.33 &  $+$0.18  &  $+$0.23 &  $+$0.25  & $-$0.79 & $+$0.12 &  $+$0.21 \\
2M17341969$-$3905457 & 16.09  &  17.87   &  14.86   &   85     & $+$1.55       &  $-$2.19   &   $-$3.30  & 4099 &  1.19 & 2.14 & $-$0.59 & $+$1.23 &  $-$0.15  &  $+$0.32 & $+$0.54 &  $+$0.36 &  $+$0.20  &  $+$0.27 &  $+$0.25  & $-$0.83 & $+$0.10 &  ...     \\
2M17342177$-$3906173 & 16.34  &  18.13   &  15.09   &   84     & $+$5.04       &  $-$2.18   &   $-$3.01 & 4166 &  1.31 & 2.10 & $-$0.35 & $+$0.92 &   ...     &  $+$0.31 & $+$0.50 &  $+$0.37 &  $+$0.11  &  $+$0.27 &  $+$0.26  & $-$0.78 & $+$0.07 &  ...     \\
2M17342943$-$3902500 & 16.22  &  17.99   &  14.97   &   77     & $+$1.26       &  $-$1.98   &   $-$3.28  & 4136 &  1.26 & 2.01 & $-$0.60 & $+$1.24 &  $-$0.12  &  $+$0.32 & $+$0.73 &  $+$0.42 &  $+$0.05  &  $+$0.26 &  $+$0.26  & $-$0.76 & $+$0.02 &  $+$0.14 \\
2M17343521$-$3903091 & 16.31  &  17.99   &  15.08   &   75     & $+$5.35       &  $-$1.98   &   $-$2.97   & 4174 &  1.33 & 2.11 & $-$0.97 & $+$1.90 &  $-$0.16  &  $+$0.22 & $+$0.18 &  $+$0.43 &  $+$0.01  &  $-$0.01 &  $+$0.55  & $-$0.87 & $-$0.01 &  ...     \\
2M17342736$-$3905102 &      17.50  &  18.88   &  16.15   &   29     & $+$4.06       &  $-$1.92   &   $-$3.09 &   ...    & ...      &  ...    &  ...     &  ...     &   ...   &     ...  &     ...  & ...     &    ...  &  ...     & ...     &   ...   &  ...    & ...      \\
\hline
{ Mean} &       &     &     &       &       &     &    &      &       &                &  {\bf  $-$0.43 }    &  {\bf   $+$0.99}    &   {\bf   $+$0.03}   &     {\bf   $+$0.30}  &  {\bf  $+$0.38}   & {\bf   $+$0.37}     &    {\bf  $+$0.21 } &  {\bf  $+$0.22 }     &  {\bf   $+$0.28}   &  {\bf  $-$0.80 }    &  {\bf $+$0.07  }   & {\bf  $+$0.09 }    \\
\hline
$1\sigma$&   &      &     &       &        &     &    &      &       &      &  $0.22$      &   $0.33$   &  $0.16$   &     $0.06$ &   $0.17$   &  $0.04$  &   $0.11$   &  $0.07$   &  $0.06$  &  $0.04$ & $0.06$   &  $0.07$\\
 	\hline 	
{\bf Spectroscopy} & & & & & & & & & & & & & & & & & & & & \\
\hline 	
2M17342588$-$3901406 &    14.69  &  16.99   &  13.35   &  234     & $+$3.63       &  $-$2.25   &   $-$3.22   &   3627    & 0.66      &  2.32   &  $+$0.07 &  $+$0.58 & $+$0.32 &  $+$0.17 &     $+$0.24  & $+$0.37 & $+$0.21 &  $+$0.18 & $+$0.27 & $-$0.84 & $+$0.05 & $+$0.07  \\
2M17341922$-$3906052 &    15.42  &  17.41   &  14.14   &  160     & $+$3.73       &  $-$2.12   &   $-$3.11   &   3783    & 0.97      &  1.51   &  $-$0.11 &  $+$0.15 & $+$0.22 &  $+$0.32 &  $+$0.08 & $+$0.38 & $+$0.12 &  $+$0.18 & $+$0.20 & $-$0.60 & $-$0.12 & $+$0.01  \\
2M17342921$-$3904514 &    15.49  &  17.38   &  14.12   &  158     & $+$2.16       &  $-$1.98   &   $-$3.39   &   4055    & 1.50      &  2.55   &  $-$0.26 &  $+$1.30 & $+$0.25 &  $+$0.13 &  $+$0.53 & $+$0.33 & $+$0.37 &  $+$0.33 & $+$0.53 & $-$0.78 & $+$0.16 & $+$0.43  \\
2M17343616$-$3903344 &    15.48  &  17.41   &  14.22   &  145     & $-$1.28       &  $-$2.34   &   $-$3.21   &   3983    & 1.44      &  2.34   &  $-$0.34 &  $+$1.22 & $+$0.22 &  $+$0.12 &  $+$0.24 & $+$0.38 & $+$0.25 &  $+$0.18 & $+$0.40 & $-$0.78 & $+$0.12 & $+$0.34  \\
2M17343025$-$3903190 &    15.48  &  17.49   &  14.16   &  144     & $+$0.63       &  $-$2.10   &   $-$3.30   &   4120    & 1.60      &  2.52   &  $-$0.51 &  $+$1.46 & $+$0.18 &  $+$0.14 &  $+$0.41 & $+$0.33 & $+$0.29 &  $+$0.33 & $+$0.54 & $-$0.72 & $+$0.15 & $+$0.55  \\
2M17342693$-$3904060 &    16.22  &  17.79   &  14.69   &  105     & $-$2.79       &  $-$2.46   &   $-$3.64   &   4146    & 1.61      &  2.25   &  $+$0.07 &  $+$0.73 & $+$0.28 &  $+$0.13 &  $+$0.17 & $+$0.39 & $+$0.12 &  $+$0.12 & $+$0.20 & $-$0.85 & $+$0.15 & $+$0.23  \\
2M17342541$-$3902338 &    15.90  &  17.60   &  14.69   &   88     & $-$2.61       &  $-$2.07   &   $-$3.31   &   4282    & 1.94      &  2.22   &  $-$0.23 &  $+$1.27 & $+$0.20 &  $+$0.12 &  $+$0.44 & $+$0.38 & $+$0.20 &  $+$0.32 & $+$0.43 & $-$0.74 & $+$0.15 & $+$0.40  \\
2M17342767$-$3903405 &    16.29  &  17.96   &  14.95   &   87     & $+$4.21       &  $-$2.41   &   $-$3.16   &   4311    & 2.05      &  2.06   &  $-$0.25 &  $+$1.22 & $+$0.17 &  $+$0.14 &  $+$0.39 & $+$0.31 & $+$0.19 &  $+$0.23 & $+$0.46 & $-$0.70 & $+$0.15 & $+$0.46  \\
2M17341969$-$3905457 &    16.09  &  17.87   &  14.86   &   85     & $+$1.55       &  $-$2.19   &   $-$3.30   &   4366    & 2.16      &  2.09   &  $-$0.01 &  $+$0.89 & $+$0.20 &  $+$0.12 &  $+$0.67 & $+$0.34 & $+$0.26 &  $+$0.31 & $+$0.48 & $-$0.73 & $+$0.17 & ...      \\
2M17342177$-$3906173 &    16.34  &  18.13   &  15.09   &   84     & $+$5.04       &  $-$2.18   &   $-$3.01   &   4361    & 2.14      &  2.01   &  $-$0.09 &  $+$0.78 &    ...  &  $+$0.09 &  $+$0.39 & $+$0.37 & $+$0.15 &  $+$0.28 & $+$0.48 & $-$0.68 & $+$0.15 & ...      \\
2M17342943$-$3902500 &    16.22  &  17.99   &  14.97   &   77     & $+$1.26       &  $-$1.98   &   $-$3.28   &   4363    & 2.07      &  1.90   &  $-$0.13 &  $+$1.05 & $+$0.18 &  $+$0.08 &  $+$0.69 & $+$0.37 & $+$0.08 &  $+$0.29 & $+$0.53 & $-$0.67 & $+$0.07 & $+$0.39  \\
2M17343521$-$3903091 &    16.31  &  17.99   &  15.08   &   75     & $+$5.35       &  $-$1.98   &   $-$2.97   &   4633    & 2.05      &  2.50   &  $-$0.69 &  $+$1.72 & $+$0.45 &  $+$0.23 &  $+$0.35 & $+$0.35 & $+$0.09 &  $+$0.07 & $+$0.26 & $-$0.68 & $+$0.02 & ...      \\
2M17342736$-$3905102 &    17.50  &  18.88   &  16.15   &   29     & $+$4.06       &  $-$1.92   &   $-$3.09   &   ...    & ...      &  ...    &  ...     &  ...     &   ...   &     ...  &     ...  & ...     &    ...  &  ...     & ...     &   ...   &  ...    & ...      \\
\hline
{ Mean} &       &     &     &       &  {\bf $+$1.92}      &     &    &      &       &                &  {\bf  $-$0.21 }    &  {\bf $+$1.03}    &   {\bf  $+$0.25 }  &     {\bf  $+$0.15}  &  {\bf  $+$0.39}  & {\bf  $+$0.36 }    &    {\bf  $+$0.20 } &  {\bf  $+$0.24 }    &  {\bf  $+$0.40  }   &  {\bf   $-$0.73}   &  {\bf  $+$0.11 }   & {\bf  $+$0.32 }     \\
\hline
$1\sigma$&   &      &     &       &     2.84   &     &    &      &       &      &  $0.20$      &  $0.32$    &   $0.06$   &     $0.04$  &    $0.17$  & $0.02$    &    $0.08$  &  $0.08$    & $0.14$   &   $0.06$  &  $0.06$    & $0.17$   \\
 	\hline 	
{\bf Typical uncertainties} & & & & & & & & & & & & & & & & & & & & \\
\hline 	
2M17342588$-$3901406 &    14.69  &  16.99   &  13.35   &  234     & $+$3.63       &  $-$2.25   &   $-$3.22   &   3627    & 0.66      &  2.32   &  0.29 &  0.15 & 0.22 &  0.13 &  0.12 &  0.16 &  0.16 &  0.13 & 0.12 & 0.19 & 0.18 & 0.12 \\
2M17341922$-$3906052 &    15.42  &  17.41   &  14.14   &  160     & $+$3.73       &  $-$2.12   &   $-$3.11   &   3783    & 0.97      &  1.51   &   0.19 &  0.21 & 0.22 &  0.16 &  0.20 &  0.05 &  0.07 &  0.12 & 0.23 & 0.14 & 0.12 & 0.13 \\
2M17342921$-$3904514 &    15.49  &  17.38   &  14.12   &  158     & $+$2.16       &  $-$1.98   &   $-$3.39   &   4055    & 1.50      &  2.55   &  0.15 &  0.17 & 0.18 &  0.10 &  0.12 &  0.12 &  0.12 &  0.11 & 0.16 & 0.15 & 0.18 & 0.07 \\
2M17343616$-$3903344 &    15.48  &  17.41   &  14.22   &  145     & $-$1.28       &  $-$2.34   &   $-$3.21   &   3983    & 1.44      &  2.34   &  0.14 &  0.21 & 0.18 &  0.09 &  0.13 &  0.11 &  0.07 &  0.09 & 0.18 & 0.12 & 0.08 & 0.10 \\
2M17343025$-$3903190 &    15.48  &  17.49   &  14.16   &  144     & $+$0.63       &  $-$2.10   &   $-$3.30   &   4120    & 1.60      &  2.52   & 0.14 &  0.16 & 0.19 &  0.08 &  0.14 &  0.11 &  0.06 &  0.09 & 0.14 & 0.13 & 0.14 & 0.11 \\
2M17342693$-$3904060 &    16.22  &  17.79   &  14.69   &  105     & $-$2.79       &  $-$2.46   &   $-$3.64   &   4146    & 1.61      &  2.25   &  0.14 &  0.18 & 0.12 &  0.10 &  0.14 &  0.10 &  0.06 &  0.06 & 0.19 & 0.11 & 0.09 & 0.06 \\
2M17342541$-$3902338 &    15.90  &  17.60   &  14.69   &   88     & $-$2.61       &  $-$2.07   &   $-$3.31   &   4282    & 1.94      &  2.22   &  0.15 &  0.26 & 0.19 &  0.12 &  0.11 &  0.15 &  0.07 &  0.12 & 0.18 & 0.11 & 0.13 & 0.08 \\
2M17342767$-$3903405 &    16.29  &  17.96   &  14.95   &   87     & $+$4.21       &  $-$2.41   &   $-$3.16   &   4311    & 2.05      &  2.06   &  0.07 &  0.15 & 0.16 &  0.11 &  0.07 &  0.08 &  0.21 &  0.07 & 0.13 & 0.10 & 0.10 & 0.10 \\
2M17341969$-$3905457 &    16.09  &  17.87   &  14.86   &   85     & $+$1.55       &  $-$2.19   &   $-$3.30   &   4366    & 2.16      &  2.09   &  0.21 &  0.26 & 0.17 &  0.11 &  0.19 &  0.11 &  0.06 &  0.10 & 0.17 & 0.13 & 0.10 & ...  \\
2M17342177$-$3906173 &    16.34  &  18.13   &  15.09   &   84     & $+$5.04       &  $-$2.18   &   $-$3.01   &   4361    & 2.14      &  2.01   &  0.04 &  0.19 & ...  &  0.11 &  0.10 &  0.12 &  0.06 &  0.09 & 0.15 & 0.11 & 0.09 & ...  \\
2M17342943$-$3902500 &    16.22  &  17.99   &  14.97   &   77     & $+$1.26       &  $-$1.98   &   $-$3.28   &   4363    & 2.07      &  1.90   & 0.24 &  0.27 & 0.19 &  0.10 &  0.12 &  0.10 &  0.11 &  0.11 & 0.19 & 0.14 & 0.10 & 0.12 \\
2M17343521$-$3903091 &    16.31  &  17.99   &  15.08   &   75     & $+$5.35       &  $-$1.98   &   $-$2.97   &   4633    & 2.05      &  2.50   & 0.20 &  0.15 & 0.26 &  0.08 &  0.06 &  0.13 &  0.07 &  0.03 & 0.13 & 0.12 & 0.11 & ...  \\
2M17342736$-$3905102 &    17.50  &  18.88   &  16.15   &   29     & $+$4.06       &  $-$1.92   &   $-$3.09   &   ...    & ...      &  ...    &  ...     &  ...     &   ...   &     ...  &     ...  & ...     &    ...  &  ...     & ...     &   ...   &  ...    & ...      \\
 	\enddata
 		\tablecomments{\bf The first part of the table show the resulting elemental abundance determined by adopting atmospheric parameters from photometry, while the second part from the spectroscopy (see Section \ref{section4}). The Typical uncertainties were computed at the same manner as described in \citet{Fernandez-Trincado2019}. $1\sigma$ is defined as (84$^{\rm th}$ - 16$^{\rm th}$)/2.}			
 \end{deluxetable*}\label{Table1}	
 %\end{rotatetable}  
  
\acknowledgments
%We thank the anonymous referee for helpful comments that greatly improved the paper. 

T.C.B. acknowledges partial support for this work from grant PHY 14-30152: Physics Frontier Center / JINA Center for the Evolution of the Elements (JINA-CEE), awarded by the US National Science Foundation. B.B. acknowledges grants from FAPESP, CNPq and CAPES - Financial code 001. D.G. acknowledges financial support from the Direcci\'on de Investigaci\'on y Desarrollo de la Universidad de La Serena through the Programa de Incentivo a la Investigaci\'on de Acad\'emicos (PIA-DIDULS). D.G. and D.M. gratefully acknowledge support from the Chilean Centro de Excelencia en Astrof\'isica y Tecnolog\'ias Afines (CATA) BASAL grant AFB-170002. 

Funding for the Sloan Digital Sky Survey IV has been provided by the Alfred P. Sloan Foundation, the U.S. Department of Energy Office of Science, and the Participating Institutions. SDSS-IV acknowledges support and resources from the Center for High-Performance Computing at the University of Utah. The SDSS website is www.sdss.org.

This work has made use of data from the European Space Agency (ESA) mission \textit{Gaia} (\url{http://www.cosmos.esa.int/gaia}), processed by the \textit{Gaia} Data Processing and Analysis Consortium (DPAC, \url{http://www.cosmos.esa.int/web/gaia/dpac/consortium}). Funding for the DPAC has been provided by national institutions, in particular the institutions participating in the \textit{Gaia} Multilateral Agreement.	

%\bibliographystyle{apj}{}
%\bibliography{references}

\end{document}